\documentclass[twocolumn,showpacs,amsmath,amssymb,prl]{revtex4}

\usepackage{graphicx}
\usepackage{dcolumn}
\usepackage{bm}

\begin{document}


\title{Sign Changes of Intrinsic Spin Hall Effect in Semiconductors
and Simple Metals: \\ First-Principles Calculations}


\author{Y. Yao$^{1,2}$, and Z. Fang$^{1,2}$}

\affiliation{ $^1$Institute of Physics, Chinese Academy of Sciences,
Beijing 100080, China \\
$^2$ Beijing National Laboratory for Condensed Matter Physics,
Beijing 100080, China \\
}

\date{\today}

\begin{abstract}
First-principles calculations are applied to study spin Hall effect in
semiconductors and simple metals.  We found that intrinsic spin Hall
conductivity (ISHC) in realistic materials shows rich sign changes,
which may be used to distinguish the effect from the extrinsic
one. The calculated ISHC in n-doped GaAs can be well compared with
experiment, and it differs from the sign obtained from the extrinsic
effect. On the other hand, the ISHC in W and Au, which shows opposite
sign respectively, is robust and not sensitive to the disorder.
\end{abstract}


\maketitle

The existence of Berry phase in systems with spin-orbit coupling (SOC)
can act as gauge field in the momentum space, which in turn affects
the transport behavior of electrons in real space, and produces the
fascinating new phenomena in solid crystals.  One typical example is
the {\it intrinsic} spin-Hall effect (ISHE) proposed
recently~\cite{Zhang1,Niu2}. The ISHE is an effect that, for
non-magnetic materials with SOC, a transverse pure spin transport can
be induced by external electric-field in the absence of magnetic field
(even at room temperature). It is distinguished from the {\it
extrinsic} spin-Hall effect (ESHE), which is due to impurity
scattering~\cite{ESHE}. The obvious advantages of ISHE, especially for
the field of spintronics, have stimulated extensive studies recently,
both theoretically~\cite{Rashba,Loss,Shen,Niu3,Sini,Inoue,OV,
SF,Sheng,Shch,Chao,Khodas,Insulator,Guo} and
experimentally~\cite{exp1,exp2}. Up to now, most of the theoretical
studies were done based on certain model Hamiltonians, like Luttinger,
Rashba, or Dresselhauss Hamiltonians. These studies provide pictures
for the understanding of deep physics (such as the effects of vertex
correction~\cite{vertex}), but on the other hand, neglect the band
details, which could be very important (as will be addressed in the
present paper) due to the topological nature of ISHE. Two experimental
evidences have been provided for the existence of spin Hall effect
(SHE)~\cite{exp1,exp2}. The SHE on the 2D hole gas~\cite{exp1} is
likely of the intrinsic origin, however, the intrinsic or the
extrinsic origin of the SHE in the 3D electron film~\cite{exp2} is
still under debate~\cite{extri,zhang3}.  In order to have close
comparison between theory and experiment, parameter-free
considerations including all band details are highly desirable, and
will be the main focus of this paper. In particular, our calculations
make reliable predictions on the sign of the ISHE, which in some cases
differs from the sign obtained from the ESHE. This qualitative
difference can be used to determine the origin of the effect.

In this letter, we will consider realistic materials by using
first-principles calculations to study the ISHE in various systems,
including semiconductors (Si and GaAs), and simple metals (W and
Au). The main difficulty of such study comes from the accurate
evaluation of Berry curvature. We have recently developed a technique
to evaluate such property accurately, and have applied it to the
calculation of the anomalous Hall effect (AHE) in Fe and
SrRuO$_3$~\cite{Fang1,Yao}. Here we apply it to study the
ISHE. Besides the quantitative evaluation, we will concentrate on the
sensitivity of ISHE to band details and the rich sign changes of ISHE
in various materials, which will provide strong support for future
experiments to identify the ISHE.

First-principles calculations have been done based on standard density
functional theory using accurate FLAPW (full potential linearized
augmented plane wave) method, in which the relativistic SOC has been
treated fully self-consistently. The exchange-correlation potential
was treated by the generalized gradient approximation (GGA), whose
validity for the systems we consider here has been shown by many other
studies. The underestimated band gap in GGA for semiconductors does
not produce problems for our purpose here, because the concerns are
for the d.c. limit ($\omega$=0). Accurate {\bf k}-point integration
has been done by tetrahedron method or adaptive mesh
refinement~\cite{Yao}. The convergence of calculated results with
respect to the number of {\bf k}-point has been carefully checked. In
general, the number of {\bf k}-point required to achieve accuracy of
5\% is around 1,000,000 in the Brillouin Zones (BZ). For all the
calculations presented here, experimental lattice parameters are used.

Suppose the external electric field is applied along the $y$
direction, then the linear response of the spin ($\sigma_z$) current
along the $x$ direction can be obtained from the Kubo formula by
evaluating the spin Hall conductivity tensor,
\begin{eqnarray}
\sigma _{xy}(\omega) &=&\frac{e}{\hbar} \int_{V_G}\frac{d^3{k}}{(2\pi )^3}
{\bf \Omega}({\bf k})
\end{eqnarray}
\begin{eqnarray}
{\bf \Omega}(\mathbf{k}) &=& \sum_nf_{nk}{\bf \Omega}_n({\bf k})
\nonumber \\ &=& \sum_nf_{nk}\sum_{n^\prime\neq
n}-\frac{2\mathrm{Im}\left\langle nk\right| j_{x}\left| n^{\prime
}k\right\rangle \left\langle n^{\prime }k\right| v_y\left|
nk\right\rangle }{\left( E_{n^{\prime }}-E_n\right) ^2-\left(
\hbar\omega +i\delta \right) ^2}. \label{eq3}
\end{eqnarray}
where $|nk>$ is the eigen wave function of Bloch state with eigen
value $E_n$ and Fermi occupation number $f_{nk}$. $v_y$ is the
velocity operator, $j_x$ is the spin current operator, which is
defined as $\frac{\hbar}{4}(\sigma_z v_x+v_x \sigma_z)$. ${\bf
\Omega}_n$ is the spin Berry phase connection of the Bloch state, and
is responsible for the anomalous transverse transport we studied. The
important point here is that all band details and SOC are
self-consistently taken into account (no adjustable parameters), no
approximation beyond linear response theory has been used. It is also
straight forward to take into account the impurity scattering effect
by allowing finite life time broadening $\delta$, and the finite
temperature effect in the Fermi distribution $f_{nk}$. For the
definition of sign, positive $\sigma_{xy}$ means that spin-up
($s_z$=1/2) component flows to the positive $x$ direction. For the
convenience of comparison, in our following discussions, we convert
spin conductivity into the unit of charge conductivity by multiplying
a factor $\frac{2|e|}{\hbar}$ to the calculated values.

\begin{figure}[!htb]
\includegraphics[scale=0.4]{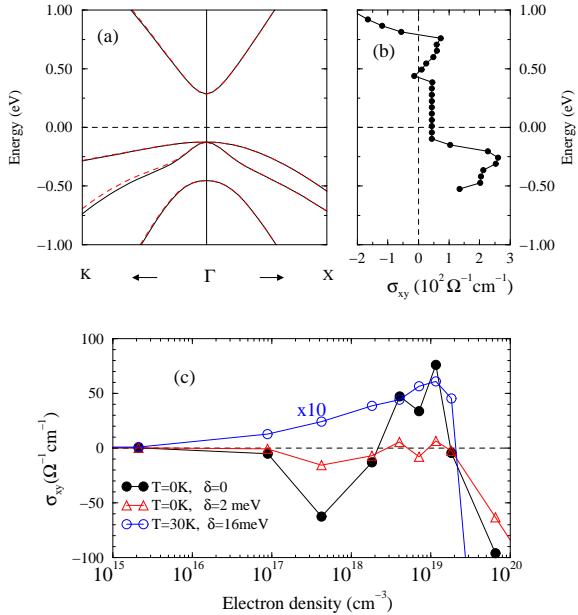}
\caption{The calculated (a) band structure; (b) spin-Hall conductivity
$\sigma_{xy}$ as function of Fermi level position for bulk GaAs.  We
define the converged Fermi level without doping as energy zero point,
and evaluate $\sigma_{xy}$ by rigidly shift the Fermi level
position. The panel (c) gives the $\sigma_{xy}$ of n-GaAs as functions
of carrier (electron) density after subtracting the part that does not
contribute to spin accumulation (see the text part for
explanation). Note the factor of 10 for the T=30K and $\delta$=16meV
curve.}
\end{figure}

\begin{figure}[!htb]
\includegraphics[scale=0.4]{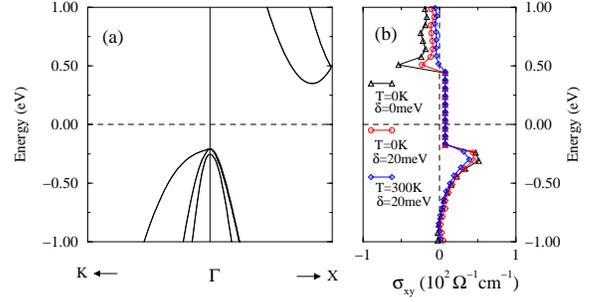}
\caption{The calculated (a) band structure; (b) spin-Hall conductivity
$\sigma_{xy}$ as function of Fermi level position for bulk Si.}
\end{figure}

{\it Semiconductors:} Fig.1 and Fig.2 show the calculated band
structures and $\sigma_{xy}$ as functions of Fermi level ($E_F$)
position for GaAs and Si respectively. In the following, we will first
concentrate on the zero-temperature and clean limit (T=0K,
$\delta$=0eV), then the effects of disorder and elevated temperature
will be addressed later.

For hole-doped GaAs (Fig.1(b)), the calculated $\sigma_{xy}$ is large,
it reaches about 300 $\Omega^{-1}cm^{-1}$, which is about the same
order of magnitude as that estimated from Luttinger
model~\cite{Zhang1}. For hole-doped Si (Fig.2(b)), on the other hand,
the maximum of $\sigma_{xy}$ is about 50 $\Omega^{-1}cm^{-1}$. This
can be understood by taking into account the fact that the strength of
SOC in Si is about 1/7.7 of that in GaAs. In both cases, the
$\sigma_{xy}$ first increases with increasing hole density, after
reaches maximum it goes down. By carefully calculate the Berry
curvature distribution $\Omega({\bf k})$, we found that the
contributions from the light hole and split-off bands are mostly
negative, which will compensate the positive contributions from the
heavy hole band, and finally suppress the $\sigma_{xy}$ for large hole
density.

For the electron-doped GaAs (Fig.1(b)), the situation is complicated
and far beyond what we understood from model analysis. The calculated
$\sigma_{xy}$ shows sign changes as electron density varies: negative
(positive) for low (high) density. Such behavior is related to a small
splitting of the conduction band due to the lack of inversion symmetry
in GaAs. For n-GaAs, the conduction band bottom has mostly $s$ orbital
character. Due to the $s$-$p$ hybridization, some $p$ characters exist
in this band, leading to the Dresselhaus type SOC in the simplified
model (therefore the vertex correction is not
important~\cite{zhang3,vertex}).

For electron-doped Si (Fig.2(b)), on the other hand, the obtained
$\sigma_{xy}$ at clean limit is quite large (although the strength of
SOC in Si is small), and has {\it negative} sign. In this case, the
conduction-band-bottom is neither around $\Gamma$ point nor $s$
orbital like.  The calculated Berry phase contributions to ISHE are
mostly related to the conduction bands around $X$ point of the
BZ. Considering the factor that the spin relaxation time in
electron-doped Si is typically much longer than that in hole-doped
case~\cite{Sarma}, our results suggest the possibility to realize ISHE
in Si, which is the most important semiconductor material.

For both GaAs and Si, the effects of temperature and disorder are
important for the electron-doped cases, while not so dramatic for
hole-doped ones. As shown in Fig.2(b) for Si, by putting
$\delta$=20meV and T=300K in our calculations, the $\sigma_{xy}$ is
significantly reduced for n-Si, while not so much for p-Si. This is
also true for GaAs as shown in Fig.1(c). The disorder and temperature
will even cause sign change for $\sigma_{xy}$ in n-GaAs as will be
discussed later in comparison with experimental results.

For the insulating GaAs and Si ($E_F$ located in the gap), however,
the calculated $\sigma_{xy}$ is non-vanishing (about 43 and 7
$\Omega^{-1}cm^{-1}$ for GaAs and Si respectively). First we have to
emphasize that this is not due to numerical error, which is four
orders of magnitude smaller. Our results can be regarded as a
generalization of the concept of ``Spin Hall
Insulator''~\cite{Insulator}. Murakami {\it et.al.} studied the spin
Hall effect in narrow-gap and zero-gap semiconductors like PbTe and
HgTe, which have ``special'' band structures (such as the inverted
light hole and heavy hole bands in HgTe), and demonstrate the
existence of spin Hall insulator. However, the results here suggest
that such ``special'' band structure is not necessary in general. In
real materials, there always exist finite hybridizations, which will
produce non-vanishing ISHE in insulators. Note that ISHE is not
quantized~\cite{Zhang1}, in qualitative difference with the
AHE. Nevertheless, we should emphasize that the existence of such ISHE
in insulator will not produce any spin accumulation due to the lack of
broken time reversal symmetry~\cite{Insulator}.

To make comparison with experimental results on n-doped
GaAs~\cite{exp2}, we show in Fig.1(c) the calculated $\sigma_{xy}$ as
functions of electron density, with the subtraction of the part that
does not contribute to the spin accumulation (the value within the
gap). Now it is very clear that the $\sigma_{xy}$ (for T=0K,
$\delta$=0meV) is negative for small doping, but change sign to be
positive for large doping. We also notice that such fluctuation of
$\sigma_{xy}$ are suppressed by introducing disorder and temperature
effects. This, on the one hand, is the nature results of topological
origin of ISHE, and on the other hand, suggests complication of ISHE
in realistic materials.  For the experimental doping density
($3\times10^{16}cm^{-3}$)~\cite{exp2}, the calculated $\sigma_{xy}$
($\delta$=0meV, T=0K) has the same sign (negative) as that obtained in
experiment, and also agrees with Ref.~\cite{zhang3}. This is in sharp
contrast with the extrinsic SHE, which has opposite sign as estimated
in Ref.~\cite{extri}.  The absolute value of calculated $\sigma_{xy}$
at clean limit is two order of magnitude larger than experimental
value (-0.005$\Omega^{-1}cm^{-1}$), however a compatible number
(-0.01$\Omega^{-1}cm^{-1}$) can be obtained by introducing a finite
lifetime broadening $\delta$=2meV as shown in Fig.1(c).  Unfortunately
experimental parameter $\delta$ for unstrained sample is
unclear~\cite{exp2}. Using the measured $\rho_{xx}$ (300 $\Omega\mu$m)
for strained sample, we estimate the $\delta$=16meV, which give
positive $\sigma_{xy}$ (T=30K, see Fig.1(c)). Nevertheless,
considering the uncertainty of experimental parameters, this issue
remains to be checked in the future.

\begin{figure}[!htb]
\includegraphics[scale=0.3]{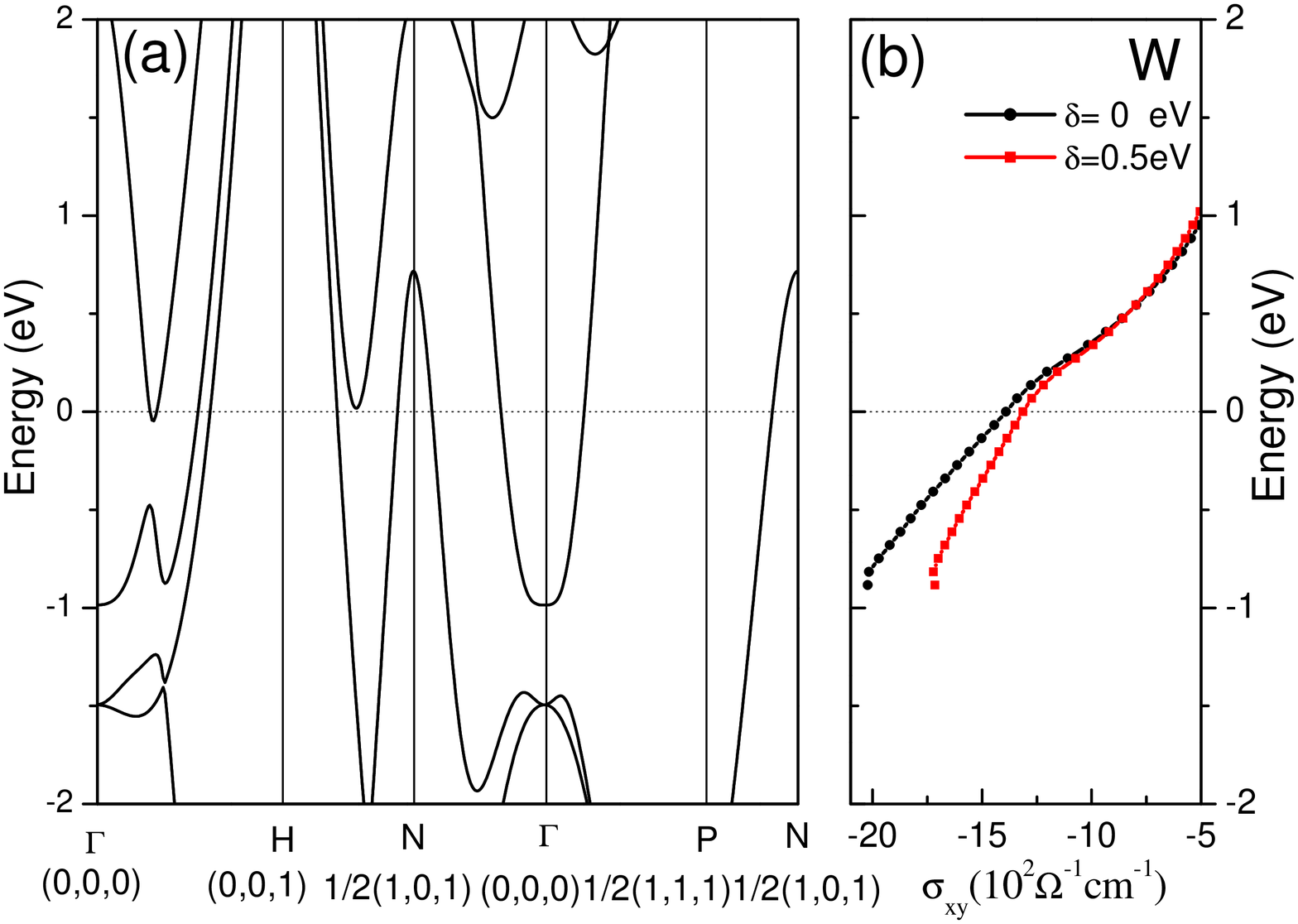}
\caption{The same notation as shown in Fig.2, but for W.}
\end{figure}

\begin{figure}[!htb]
\includegraphics[scale=0.3]{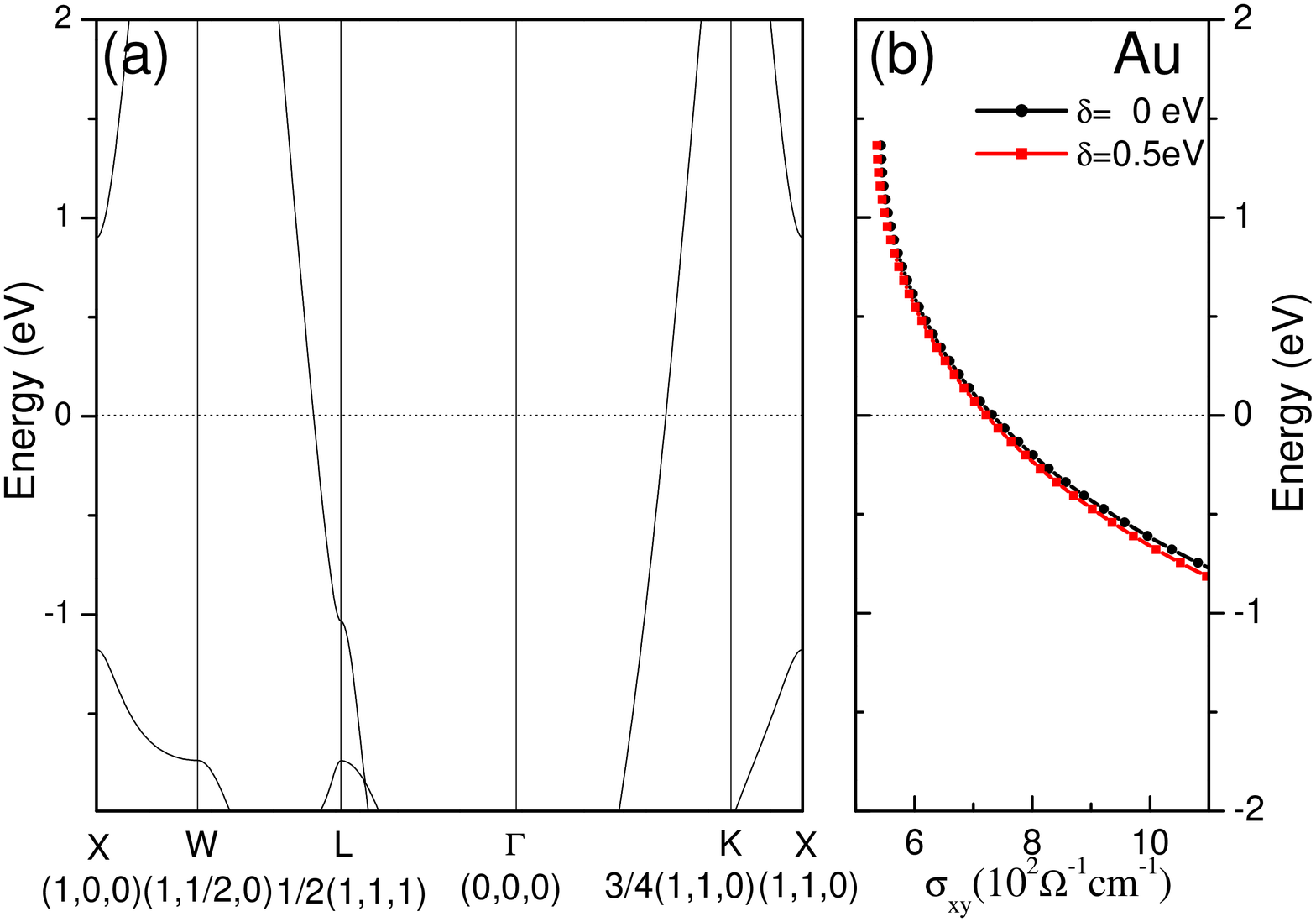}
\caption{ The same notation as shown in Fig.2, but for Au.}
\end{figure}

{\it Simple Metal:} The ISHE in simple metal has not been studied yet,
although it is not surprising to expect that ISHE exists in such
systems, due to the same mechanism. We chose elemental W and Au as
examples because of the relatively larger SOC. For W the charge
conductivity mostly comes from $5d$ states around Fermi level, while
for Au it is mostly from $6s$ states. The calculated band structures
and $\sigma_{xy}$ are shown in Fig.3 and Fig.4 for W and Au
respectively.  Besides the very strong ISHE obtained (the
$\sigma_{xy}$ can reach as high as -1390 $\Omega^{-1}cm^{-1}$ for W,
and 731 $\Omega^{-1}cm^{-1}$ for Au), we notice that $\sigma_{xy}$ is
negative in W, but positive in Au.  This again suggests the rich sign
changes of ISHE in realistic materials. What is more interesting is
that the ISHE in W and Au are robust and not sensitive to the disorder
(in opposite to GaAs or Si). By adding a very large disorder effect
($\delta$=0.5eV), the calculated $\sigma_{xy}$ only change
slightly. Given these special characters, we suggest that both W and
Au are nice candidates for future experimental examination of
ISHE. Especially for Au, where the conduction electrons have mostly
$s$ characters, relatively long spin relaxation can be
expected~\cite{Sarma}

{\it Sign Issue and Discussions:} As presented in our above results,
the ISHE shows very rich sign changes, which are independent of the
carrier type and the sign of impurity potential: (1) for Si, the sign
of ISHE is the same as ordinary Hall effect, i.e., positive for hole
doping and negative for electron doping; (2) for W and Au, however,
the sign of ISHE is opposite with their carrier type: W (Au) has hole
(electron) type conductivity but negative (positive) ISHE; (3) for
n-doped GaAs, the sign of ISHE changes with increasing doping. Such
rich sign changes are more than what we can expect from extrinsic
scattering mechanism~\cite{ESHE}. Two mechanisms, namely skew
scattering and side-jump, were mainly discussed in
literatures~\cite{Fivaz}.  For a simple discussion, we consider the
skew scattering mechanism~\cite{Fivaz}, which dominates over side-jump
contribution for the weak disorder limit. In this case, the sign of
ESHE depends on the sign of scattering potential~\cite{Fivaz}.  It is
natural to expect that the sign of skew scattering does not change
with changing impurity density. However, we predict that the sign of
ISHE can change with the same type of doping (in n-GaAs).  Such
difference may be used to distinguish ESHE from intrinsic
contributions.  We can also use simple metal W and Au for such
purpose, because opposite signs of ISHE (which has nothing to do with
scattering potential) are predicted. Nevertheless, {\it the sign issue
should be regarded as a very important aspect of SHE, and can be used
in future experiments}.

In summary, we have performed detailed studies on the ISHE for various
realistic materials based on accurate parameter-free first-principles
calculations, and predict rich sign changes of ISHE. Furthermore, we
demonstrate that the ISHE in semiconductors (GaAs and Si) is highly
sensitive to band-details and disorder, while ISHE in simple metals (W
and Au) is robust and not sensitive to disorder. The calculated ISHE
for n-GaAs can be well compared with experimental results, while the
extrinsic spin Hall contribution has opposite sign.


\end{document}